\documentclass[useAMS,usenatbib]{mnras} 
\usepackage{graphicx}
\usepackage{amsmath}
\usepackage{bm}
\usepackage{url}
\usepackage{color}
\usepackage{arydshln}
\usepackage{lscape}
\usepackage[section]{placeins}
\usepackage[normalem]{ulem}
\usepackage{lineno}
\usepackage{booktabs}

\begin{document}


\title{Effects of baryons on the gravitational redshift profile of $\Lambda$CDM halos}

\author[Zhu et al.]{\parbox{18cm}{Hongyu Zhu$^{1,2}$\thanks{E-mail:
hongyuz@andrew.cmu.edu}, Shadab Alam$^{3,1,2}$, Rupert A. C. Croft$^{1,2,4}$, Shirley Ho$^{1,2,5,6,7}$, Elena Giusarma$^{1,2,5,6}$}
\vspace{0.3cm}\\
    $^{1}$ Department of Physics, Carnegie Mellon University, 5000 Forbes Ave., Pittsburgh, PA 15213 \\
    $^{2}$ McWilliams Center for Cosmology, Carnegie Mellon University, 5000 Forbes Ave., Pittsburgh, PA 15213 \\
    $^{3}$ Institute for Astronomy, University of Edinburgh, Royal Observatory, Blackford Hill, Edinburgh, EH9 3HJ , UK\\
    $^{4}$ ASTRO-3D Center, School of Physics, University of Melbourne, Parkville, VIC 3010, Australia\\
    $^{5}$ Berkeley Center for Cosmological Physics, University of California, Berkeley, CA 94720, USA\\
    $^{6}$ Lawrence Berkeley National Laboratory (LBNL), Physics Division, Berkeley, CA 94720, USA\\
    $^{7}$ Flatiron Institute, Center for Computational Astrophysics, 162 Fifth Ave., New York, NY, USA 10010
}
    
\date{\today}
\pagerange{\pageref{firstpage}--\pageref{lastpage}}   \pubyear{2018}
\maketitle
\label{firstpage}

\begin{abstract}
Gravitational redshifts and other relativistic effects are beginning
  to be studied in the context of galaxy clustering.  Distortions
  consistent with those expected in General Relativity have been
  measured in galaxy cluster redshift profiles by Wojtak et al. and
  others and in the the cross-correlation function of galaxy
  populations by Alam et al. On scales below $\sim$20 Mpc/$h$
  simulations have shown that gravitational redshift dominates over
  other effects. However, this signal is related to the shape and
  depth of gravitational potentials, and therefore the matter density
  in galaxies and galaxy clusters that is responsible for them. We
  investigate the effects of baryonic physics on the gravitational
  redshift profiles of massive (group and cluster-sized) halos. We
  compare the profiles of different components in halos taken from the
  MassiveBlack-II cosmological hydrodynamic simulation and a dark
  matter-only version of the same simulation. We find that inclusion
  of baryons, cooling, star formation and feedback significantly
  alters the relevant inner density profiles.  These baryonic effects
  lead to overall increases in both gravitational redshifts and the
  transverse relativistic Doppler effects by up to $\sim$50\% for group size
  halos. We show how modified Navarro Frenk White 
 halo profiles can be used to parametrize these differences, and provide
relevant halo profile fits.
\end{abstract}


\section{Introduction} \label{sec:intro}

The structure of Universe  is thought to have formed from the growing modes of tiny quantum fluctuations in an initial homogeneous field. Such growing modes collapsed through gravity and formed what we call today dark matter halos. These dark matter halos provided the necessary environment for baryonic component of the matter to collapse further and form stars at the smallest level and larger structures like galaxies, collections of stars held together in gravitationally nurturing environment of dark matter halos. The baryonic component's evolution through several highly energetic events of star and galaxy formation also shapes the dark matter itself especially in the innermost regions. All such effects taken together in the structure formation paradigm are known as baryonic effects on dark matter halos. These become extremely important for studies which are sensitive to the innermost parts of the halos for example gravitational lensing \citep{dodelson2003modern, Zentner2013, Huang2018}, SZ effects, galaxy velocity bias, and gravitational redshift.


Several attempts have been made to understand and predict the nature of the inner structures of 
dark matter halos. \citet{navarro1996, navarro1997} showed that density
profiles of simulated dark matter halos have all the same shape, independent of the
halo mass, the initial density fluctuation spectrum, and the values of
the cosmological parameters, by studying the equilibrium density
profiles of dark matter halos from $N$-body models. The NFW
density profile can be expressed with the simple formula
\begin{equation}
\label{nfw_density}
\rho (r) = \frac {\rho_\mathrm{crit}(z)\delta_c} {(r/{r_s})\left(1+r/{r_s}\right)^2},\delta_c = \frac {200} 3 \frac {c^3} {\ln(1+c)- c/(1+c)}
\end{equation}
where $\rho_\mathrm{crit}(z)=3H^2(z)/8\pi G$ is the critical density
at redshift $z$, $r_s$ is the scale radius and $\delta_c$ is a
characteristic (dimensionless) density. $c=R_{200}/r_s$ is the
``concentration'' of the halo that provides a link between $r_s$ and
$\delta_c$. \citet{navarro1997} also found that massive halos are less concentrated
than smaller halos.  The very inner slope in
Eqn.~\ref{nfw_density}, $\mathrm{d} \ln \rho(r)/\mathrm{d} \ln
r|_{r\rightarrow0}=-1$. This inner slope is difficult to measure,
and very sensitive to baryonic physics, something which was not modelled in the initial dark matter halo  studies. For instance, \cite{moore1999}
used a profile with a steeper inner slope -1.5. A generalized NFW
(hereafter gNFW) density profile (e.g. \citealt{hernquist1990,zhao1996}) was proposed, being:
\begin{equation}
\label{gnfw_density}
\rho (r) = \frac {\rho_\mathrm{crit}(z)\delta_c} {(r/{r_s})^\gamma\left(1+r/{r_s}\right)^{3-\gamma}}~.
\end{equation}
In Eqn.~\ref{gnfw_density}, $\mathrm{d} \ln \rho(r)/\mathrm{d} \ln
r|_{r\rightarrow0}=-\gamma$ is an additional free parameter
(e.g. \citealt{schmidt2007}).  \citet{navarro2004} also explored a more
general model, similar to that of de Vaucouleurs \citep{de1948} in which the
logarithmic slope varies continuously with
radius. This model happened to be the same as one
 introduced by Jaan Einasto at a 1963 conference in Alma-Ata, Kazakhstan
  \citep{einasto1965}.   It is therefore referred to as ``Einasto's
model''. Einasto's  model is expressed as
\begin{equation}
\label{einasto_density}
\rho(r) = \rho_{-2}\exp\left\{-\frac 2 \alpha \left[\left(\frac r {r_{-2}}\right)^\alpha - 1\right]\right\}~.
\end{equation}
As pointed out by \cite{merritt2005}, Einasto's model has the same
slope-radius relation as Sersic's model \citep{sersic1963}
 and works well for both
projected (surface) density and (space) density profiles of
galaxies. Eqn.~\ref{einasto_density} can be also written as
$\mathrm{d} \ln \rho(r)/\mathrm{d} \ln r=-2(r/r_{-2})^\alpha$ where
$r_{-2}$ is the radius that yields a  $\rho(r)$ slope  in
log-log space of $-2$, with $\rho_{-2}$  the density at that radius. It
can be seen that Einasto's model is a generalization of a power law in
log-log space and  that the parameter $\alpha$ controls curvature of the
profile.  Einasto's model has been used to describe many types of
system including galaxies and dark matter halos and it is  often  considered a
better fit than the NFW model to  $N$-body dark matter
halo profiles \citep{merritt2006}.

Baryonic effects such as radiative cooling and feedback have been
investigated in a variety of simulations such as the OverWhelmingly
Large Simulations project (OWLS), Feedback In Realistic Environments
(FIRE), Millennium Simulation (MS), Evolution and Assembly of Galaxies
and their Environment (EAGLE) \citep{neto2007, Rudd2008,duffy2010,Velliscig2014,chan2015, schaller2015, Schaller2015b}. It has been claimed that simulated dark matter profiles
would be steeper in the inner regions than the NFW profile if the
cooling is efficient and feedback is weak since the gravity of 
the central baryons
pulls dark matter towards the center; without this, dark matter
profiles would be shallower. The slope of the inner dark
matter density profile therefore shows a strong mass dependence. To understand
the structure of the inner halo is crucial as it is not only of
major importance for efforts to detect dark matter experimentally but
also central to possible issues which have been 
identified with the $\Lambda$CDM paradigm from small scales,
such as the {\it cusp/core} problem \citep{de2010}.

Gravitational redshift is one of the major predicted effects that 
occur in  weak
gravitational fields according to the principle of equivalence from
General relativity \citep{Einstein1916}. A photon with wavelength
$\lambda$ emitted in a gravitational potential $\Phi$ and observed at
infinity has a gravitational redshift $z_\mathrm{g}=\Delta
\lambda/\lambda\approx \Delta\Phi/c^2$. The effect has been
 been measured for the
Earth’s gravity, in the Solar system and in white dwarf stars
(e.g. \citealt{pound1959,takeda2012,falcon2010}) as one of the
fundamental tests of General relativity. By assuming an analytic de
Vaucouleurs profile, \cite{cappi1995} argued that gravitational
redshifts of  central galaxies in clusters with respect to the other cluster
members would be tens of km/s in the most massive galaxy clusters,
which is feasible to detect. \citet{kim2004} used $N$-body simulations
to show that it should be
possible to detect the radial profile of  gravitational redshifts 
in galaxy clusters. \citet{wojtak2011} carried out the first
 observational measurement of 
gravitational redshifts in galaxy clusters and found consistency with
General relativity. 

\citet{croft2013} predicted a $\sim4\sigma$ signal
of gravitational redshifts from large-scale structure should be
obtainable from the full Baryonic Oscillation Spectroscopic Survey
(BOSS) by examining the distortion due to gravitational redshifts of
the cross-correlation function of two galaxy populations. 
The relativistic distortions cause a dipole in clustering which is
not present when they are not included \citep{mcDonald2009}
 Besides the gravitational redshift
effect, there exist other relativistic effects comparable in  magnitude
such as the transverse Doppler effect, light cone bias and special
relativistic beaming, as pointed out by \citet{kaiser2013} and
\cite{Giusarma2016PT}, and modeled in simulations by \citet{zhao2013},
\citet{cai2016}, \cite{Alam2016TS} and \cite{Zhu2016Nbody}. These
were included in comparisons to an observational
measurement of relativistic distortions in galaxy clustering
\cite{Alam2016Measurement}.
These effects are included to linear order in the GR perturbation
theory treatments of \citep{yoo2009,yoo2012,bonvin2014}.
 Most relevant to the current paper is
the fact that \cite{Zhu2016Nbody},  and \cite{Breton2018} find that
gravitational redshift dominates the non-linear regime, and consequently
the range of scales where signal to noise of measurements is largest.
 Here the resolution
of $N$-body simulations
 and the clustering of baryons on small scales
are crucial for  measuring and understanding
relativistic effects.

In this paper
we use the MassiveBlack-II (hereafter MBII) high
resolution hydrodynamical simulation \citep{khandai2015} to predict
the effects of baryonic physics on the gravitational redshift
of galaxies. We also compare to a dark matter-only simulation (hereafter DMO)
run with the same volume, cosmological parameters, and initial conditions
\citep{tenneti2015}. Comparing the outputs from these two enables us
to study the baryonic effects in dark matter halos.

This paper examines the small scale behavior of gravitational
redshift profiles in and around galaxies as well as the effects of
baryons on small scales. Our plan for the paper is as follows. We 
discuss the simulations we used, the approach to the sample and how we
obtain the gravitational potential from MBII and
its corresponding dark matter-only simulations in Sec.~\ref{sec:sim}. In
Sec.~\ref{sec:profile} we show the density and gravitational profiles
of massive halos from different components and simulations. The comparison between 
simulation (spatial) and observational
(projected) quantities is explored in the last part of
Sec.~\ref{sec:profile}. In Sec.~\ref{sec:fitting}, we fit analytic
models to the profiles, demonstrating the effects of baryons in a more
quantitative way. Sec.~\ref{sec:velocity} presents another
comparable relativistic effect, the transverse Doppler effect
and explores how baryonic physics affects it. We conclude
in Sec.~\ref{sec:discussion} with a summary and a discussion of our
findings.

\section{Simulations}\label{sec:sim}
The hydrodynamic simulation we use,  MBII
 \citep{khandai2015} was run using  {\small P-GADGET}, a
massively parallel TreeSPH cosmological simulation code combining a
collisionless fluid with the $N$-body method and an ideal gas by means
of smoothed particle hydrodynamics (SPH) (see \citealt{springel2005} for a predecessor public version,
{\small GADGET-2}). The cosmological parameters in MBII and DMO simulation are $\sigma_8=0.816$, $n_s=0.968$, $\Omega_\lambda=0.725$, $\Omega_m=0.275$, $\Omega_b=0.046$ (MBII) and $h=0.071$. Simulation parameters are listed in Tbl.~\ref{tbl:sim}.
\begin{table}
\centering
\caption{Simulation parameters in MBII and DMO: $N_\mathrm{part}$, $m_\mathrm{DM}$, $m_\mathrm{gas}$ and $\epsilon=1.85\,\mathrm{kpc}/h$ }
\label{tbl:sim}
\begin{tabular}{llll}
\hline
       & $N_\mathrm{part}$  &  $m_\mathrm{DM}$ [$M_\odot/h$]& $m_\mathrm{gas}$ [$M_\odot/h$] \\ \hline
MBII   &  $2\times1792^3$   &$1.1\times10^7$&$2.2\times10^6$\\ \hline
DMO    &  $1792^3$          &$1.32\times10^7$& NA \\ \hline
\end{tabular}
\end{table}
MBII is large enough to contain massive halos of mass
above $10^{14}\;M_\odot/h$ as the cubical comoving volume is
$(100\;\mathrm{Mpc}/h)^3$ \citep{khandai2015}. MBII includes the models for star formation, black hole (BH) growth and radiative cooling and heating processes. Such massive halos are
 expected to show a $\sim 10 \,km/s$ difference in gravitational
 redshift between their centers and edges.
  Furthermore, 
 a corresponding dark matter-only simulation to MBII was performed with the same volume, cosmological parameters, and initial conditions  
 \citep{tenneti2015}. Comparing the results from these two enables us to study the impact of baryonic processes on the properties of dark matter halos.

\subsection{Halos and subhalos}
As described in \citep{khandai2015, tenneti2015}, the halos are identified through the friends-of-friends (FOF) algorithms \citep{davis1985} on dark matter particles with a linking length of $b=0.2$ times the mean of particle separations. Gas and stars are associated to their closest dark matter particles. {\small SUBFIND} code \citep{springel2001} is used to find subhalos which are locally overdense, self-bound groups of particles within halos. The process starts with isolated local density peaks inside halos and expands until the overdensity reaches a certain threshold.

\subsection{Efficient Calculation of gravitational redshift potential}

\begin{figure}
\centering
	\includegraphics[width=1\columnwidth]{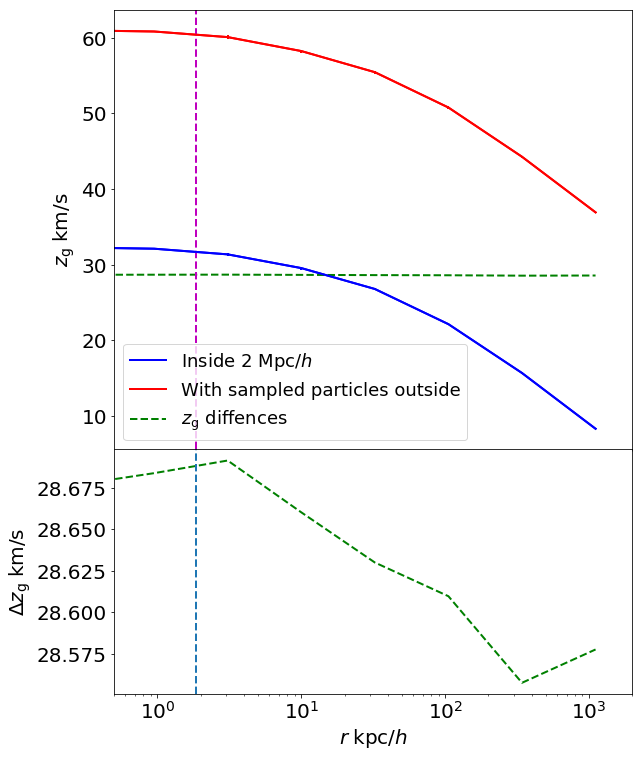}
 \caption{ A test of 
the gravitational redshift calculation:
The gravitational redshift profiles of particles within
   $2\;\mathrm{Mpc}/h$ (blue), of both inside particles and sampled
   $N_\mathrm{sample}$ particles outside (red) and their difference
   (green) in  the MBII simulation in the halo of mass $>
   7\times10^{14}\;M_\odot/h$. The center is chosen at
   the potential minimum in the halo. The bottom panel magnifies the
   difference (green) between red and blue lines in the top panel.}
 \label{fig:z_difference}
\end{figure}

Generally cosmological simulations produce a massive amount of data which are difficult to store with limited disk space. Traditionally the gravitational potential of particles are not stored as it is not needed for standard analysis. We therefore devise a method to obtain the potentials for each particle in the simulation in post-processing.
Given enough resources, the simplest method to obtain the potential is to use the same N-body code initially used to produce the simulations, loading all the particles and velocities and running for a single time step with instruction to also save the calculated potential. However, since only the potentials for the particles within selected halos are needed, and local computing resources are not able to run the full 11.5 billion particle MBII hydro simulation, we use an alternative technique. We select all particles within $2\;\mathrm{Mpc}/h$ of the halo centres since we are only interested in the gravitational redshift profile on small scales, and also randomly sample $N_\mathrm{sample}=2,000,000$ particles outside this range. For the sparse sample particles, we multiply the mass of sampling particles by a factor of $N_\mathrm{outside}/N_\mathrm{sample}$ where $N_\mathrm{outside}$ denotes the total number of particles outside $2\;\mathrm{Mpc}/h$. We calculate  gravitational redshift profiles both with and without the sampled $N_\mathrm{sample}$ particles outside and compute the difference between them (See Fig.~\ref{fig:z_difference}). The gravitational redshift profiles are quite smooth and the difference is a constant as expected, though with $<$1\% fluctuations that are barely visible. Therefore, in our further discussions about gravitational redshifts, we only use the particles inside the $2\;\mathrm{Mpc}/h$ radius since a constant in $z_g$ (i.e. potential) can be considered the same as assigning a different zero-point potential which makes no difference to the  physics.

\section[]{Density and gravitational profiles of massive halos}
The dark matter haloes and galaxies lives in three spatial dimensions but in observations we are always limited to projected profile in the plane of sky. Therefore it is important to distinguish the two and understand the impact of projection on the measurement. In the following sections we discuss both three dimensional spatial profiles and projected profiles while contrasting the two with each other.

\subsection{Spatial (deprojected) profiles} \label{sec:profile}

\begin{figure}
\centering
\includegraphics[width=\columnwidth]{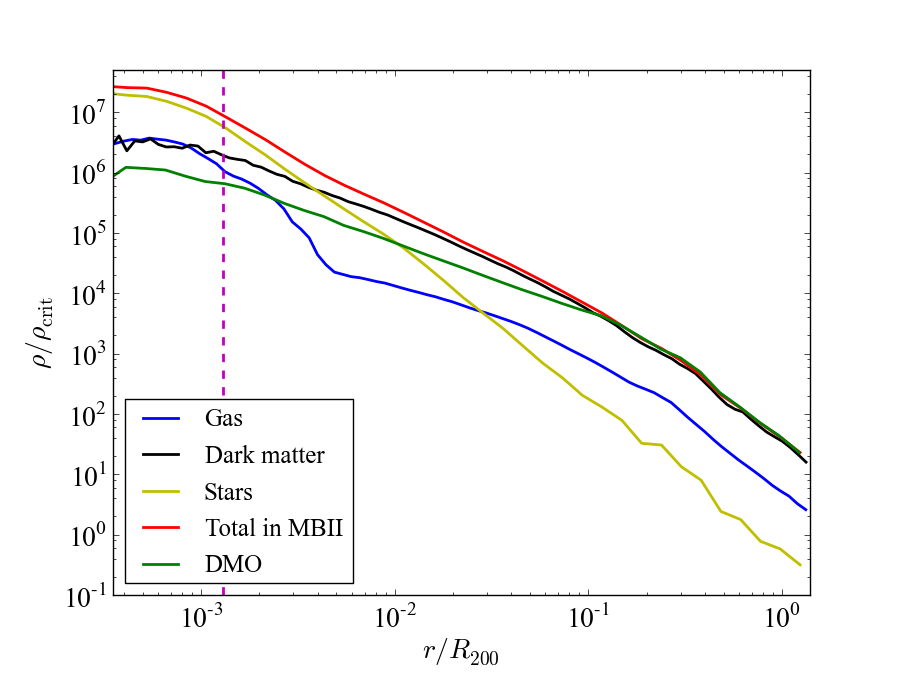}
 \caption{The density profiles of gas (blue), dark matter (black),
   stars (yellow) and total matter (red) in MBII and dark matter
   (green) in DMO simulations in a halo of mass $
   7\times10^{14}\;M_\odot/h$. The center is chosen to be at the potential
   minimum in the halo. The $r$ range plotted is from the
   gravitational softening $\epsilon = 1.85\;\mathrm{kpc}/h$ up to
   $2\;\mathrm{Mpc}/h$.}
 \label{fig:density}
\end{figure}

We concentrate on the most massive halo at $z=0.06$ in the 
MBII hydrodynamical simulation (MBII)  and
its  equivalent in  MBII dark matter-only
(DMO) simulation. The halo has a mass $
7\times10^{14}\;M_\odot/h$ and most of that mass is
in a large subhalo of mass $
6\times10^{14}\;M_\odot/h$. There are slight differences in mass and
positions of the halo between the two simulations. The profiles
are calculated using the potential minima  of this halo as the
center and the small offset between MBII and DMO is 
accounted for. We first want to understand the scale dependence of dark matter density and which parts of halos are dominated by different components. We show the angular averaged density profile from the center of halos in Figure~\ref{fig:density}. It can be seen that the
distributions of total matter in the DMO and MBII simulations 
are nearly identical on large scales, showing that baryonic
effects are negligible beyond 0.1 $r_{200}$. For
MBII, dark matter is the major component for radii larger than
$\sim 0.005R_{200}$. Stellar matter is the dominant component for the
inner radii (below $\sim 0.005R_{200}$) and exhibits a steeper profile within
this radius. The gas is concentrated near the center and becomes negligible at
radii larger than $\sim 0.005R_{200}$. For DMO, the dark matter
profile is much flatter close to the center compared to the dark
matter profile in MBII, lacking the contraction caused by the gravity of 
a core of stars and dense cooling gas.

\begin{figure}
  \centering
    \includegraphics[width=1\columnwidth]{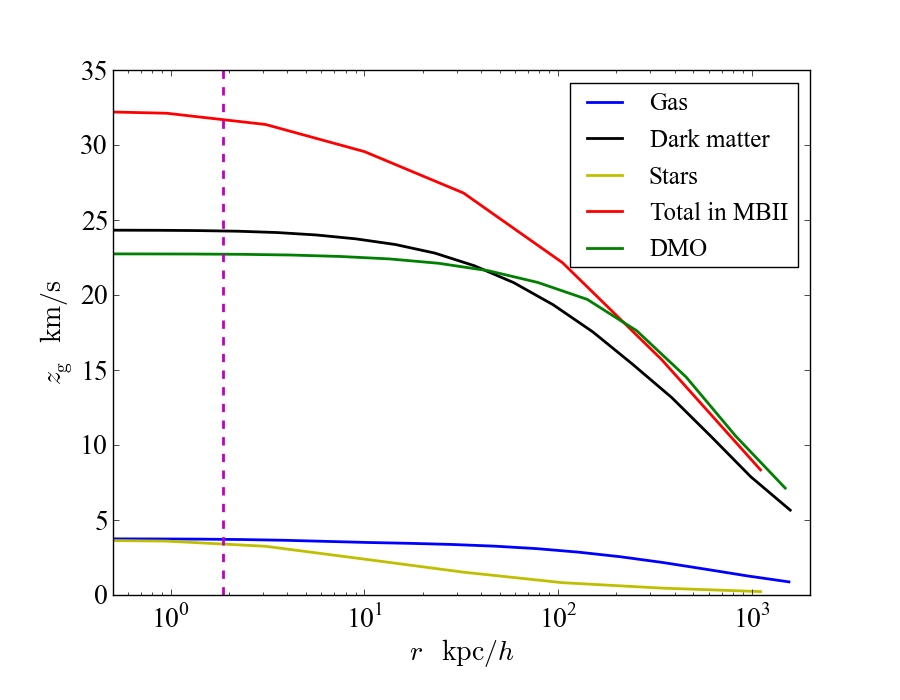}
     \caption{The gravitational redshift profiles of gas (blue), dark
       matter (black), stars (yellow) and total matter (red) in MBII
       and dark matter (green) in the DMO simulations in the halo of mass
       $ 7\times10^{14}\;M_\odot/h$ in log-log space. The center is
       chosen at the potential minimum of the halo. The distance
scale  $r$
       spans the gravitational softening $\epsilon =
       1.85\;\mathrm{kpc}/h$ to $2\;\mathrm{Mpc}/h$.}
     \label{fig:z_types}
\end{figure}

Fig.~\ref{fig:z_types} shows the contributions from gas, dark matter
and stars to the gravitational redshift profile of the same halo. The 
differences between the center and edges are of the order of
$\sim10\,\mathrm{km/s}$ as expected. Gas and stellar mass contribute
$\sim$ 1 order of magnitude less than dark matter because dark matter
is similarly less abundant in mass. Though stellar mass
is more concentrated in the center, it quickly becomes sparser at 
distances larger than $\sim 0.005R_{200}$. As a result, it still
makes the smallest contribution to the gravitational redshift around the
center and its contribution decays rapidly with increasing radius. We
can also see in Fig.~\ref{fig:z_types} that the existence of baryons
steepens the gravitational redshift profile as the dark matter has
been pulled inward during cooling. We also investigate other 8 massive
halos and find there is in all cases a 20\% - 50\% increment in $z_g$ in the
centers for halos simulated with hydrodynamics relative to the DMO case.
 This size of this fractional increment depends on halo mass,
and presence of substructures.

\begin{figure*}
  \centering
    \includegraphics[width=1\textwidth]{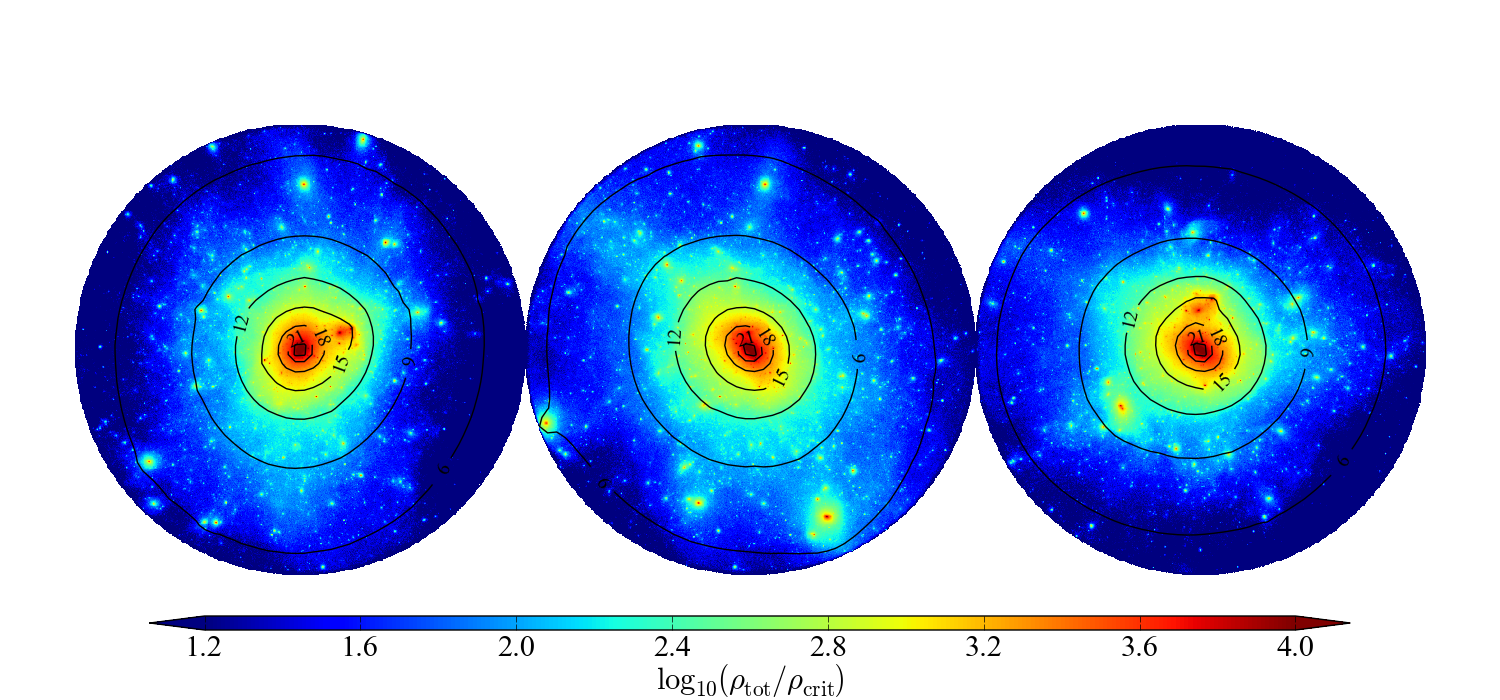}
     \caption{2D plots showing the density of all types of particles (gas,
       dark matter, stars) in the MBII simulation in a slice of
thickness  $\pm
       400\;\mathrm{kpc}/h$ diameter $4\;\mathrm{Mpc}/h$
 centered on the gravitaional potential minimum.  We show 3 different
orthogonal       planes. The colormaps show the density and the
       contours represent the gravitational redshift. The subplots are
       the slices  in the $X-Y$, $X-Z$ and $Y-Z$ planes respectively
       from left to right.}
     \label{fig:sliceplot}
\end{figure*}

Fig.~\ref{fig:sliceplot} shows the density distribution in the 
most massive halo as well as contours of gravitational redshift. The halo 
contains substantial substructure. The
gravitational redshift contours are much smoother than the mass
distribution.

\subsection{Surface (projected) profiles}
\begin{figure}
  \centering
    \includegraphics[width=1\columnwidth]{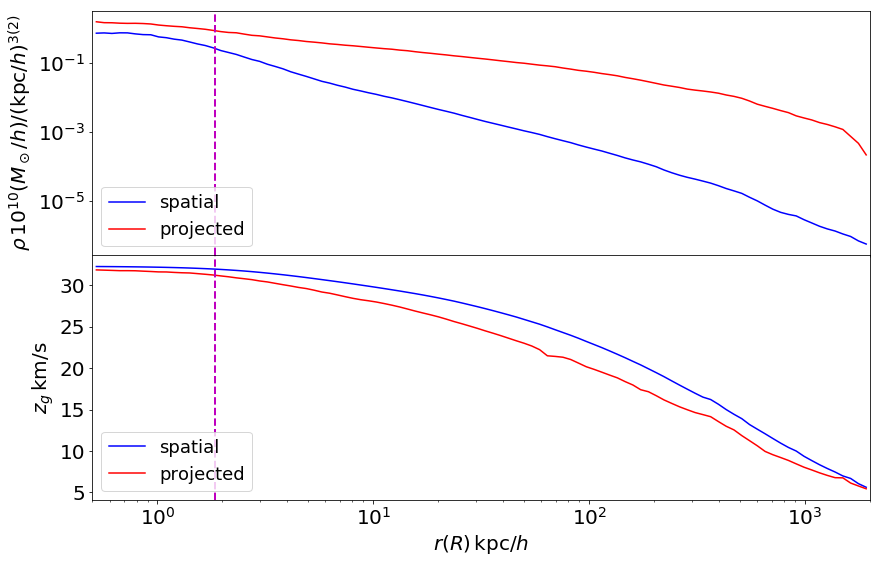}
     \caption{
The density (top panel) and gravitational
       redshift (bottom panel) profiles of total matter in MBII in the
       halo of mass $ 7\times10^{14}\;M_\odot/h$. The center is
       chosen at the potential minimum. In each panel,
       the spatial (deprojected) profile is shown in blue while the
       surface (projected) profile is shown in red. The distance
      scale  $r$ spans the  gravitational softening $\epsilon
       = 1.85\;\mathrm{kpc}/h$ to $2\;\mathrm{Mpc}/h$.}
     \label{fig:projected}
\end{figure}

In order to relate our simulation results to quantities that
are directly  measurable from  observations, we turn to projected profiles.
 We investigate the spatial (deprojected) and surface
(projected) profiles of both the  density and the gravitational redshift in
Fig.~\ref{fig:projected}. 
We also choose the center of our profiles to be the potential minimum of the halo.
To estimate the density profiles, we accumulate the
mass in each bin and divide it by the bin area or volume as appropriate
  ($4\pi r^2\Delta r/3$) for spherical bins and ($2\pi R\Delta
R$) for cylindrical bins; this leads to a difference in the units, (see Eqn.~\ref{eq:dpro}):
\begin{subequations}\label{eq:dpro}
    \begin{align}
    & \rho_{k,\mathrm{spatial}} = \frac{\sum_{r_i \geq r_k-\Delta r/2}^{<r_k+\Delta r/2}m_i}{4\pi \left[\left(r_k+\Delta r/2\right)^3-\left(r_k-\Delta r/2\right)^3\right]}, \\
    & \rho_{k,\mathrm{projected}} = \frac{\sum_{R_i \geq R_k-\Delta R/2}^{<R_k+\Delta R/2}m_i}{2\pi \left[\left(R_k+\Delta R/2\right)^2-\left(R_k-\Delta R/2\right)^2\right]},   
    \end{align}
\end{subequations}
where $k$ indicate the radial bin number, $m_i$ is the mass of the $i$-th particle. $r_k$ and $R_k$ are the distances to the center of the $k$-th bin and $\Delta r$ and $\Delta R$ are bin sizes.

To compute the gravitational redshift profiles we use the mass weighted
gravitational redshift in spherical or cylindrical bins (see Eqn.~\ref{eq:zpro}):
\begin{subequations}\label{eq:zpro}
    \begin{align}
     & z_{g,\mathrm{spatial}} = \frac{\sum_{r_i \geq r_k-\Delta r/2}^{<r_k+\Delta r/2}z_i}{\sum_i I({r_k-\Delta r/2 \leq r_i<r_k+\Delta r/2})},\\
     & z_{g,\mathrm{projected}} = \frac{\sum\limits_{R_i \geq R_k-\Delta R/2}^{<R_k+\Delta R/2}z_i}{\sum_i I({R_k-\Delta R/2 \leq R_i<R_k+\Delta r/2})},
     \end{align}
\end{subequations}
where $I(\cdot)$ is an indicator function.

We find that for density profiles the surface profile is much flatter than the
spatial one while the trend is completely different for gravitational
redshift profiles (projected being almost the same as spatial, and even slightly steeper). The reason for this  is that spatial and surface
density profiles are normalized by different bin sizes, thus the
profile has different dependencies on $r$ or $R$. In
general, for quantities such as gravitational redshift,  the surface
profile $F(R)$ can be derived from the spatial profile $f(r)$ using
Abel transform, see \cite{abel1826auflosung}
\begin{equation}
F(R) = 2
\int\limits_{R}^{\infty}\frac{f(r)r\mathrm{d}r}{\sqrt{r^2-R^2}}, \end{equation}
where $r$ and $R$ are spatial and projected distance to the center
respectively. We can learn from the above equation that the projected profile
 averages quantities along the line-of-sight so that more edge
information is brought in. Thus, in the bottom panel of
Fig.~\ref{fig:projected}, the
projected profile at $R$ lies in between  the spatial profile at
$r=R$ and the outskirts.

\section{Analytic models of density and gravitational redshift profiles of
  massive halos}
\label{sec:fitting}

As mentioned in Sec.~\ref{sec:intro}, the dark matter density profile of
a dark matter halo can be described by either NFW (See
Eqn.~\ref{nfw_density}), gNFW (See Eqn.~\ref{gnfw_density}) or
Einasto (See Eqn.~\ref{einasto_density}) models. The NFW profile is the simplest and the most commonly used dark matter profile. However, it is not able to control the slope of the inner profile, which is most affected by baryonic effects. Two other profiles, gNFW and Einasto are introduced to better model the dark matter halo and potentially can be used to describe the baryonic effects, by parametrizing the inner slope as $\mathrm{d} \ln \rho(r)/\mathrm{d} \ln
r|_{r\rightarrow0}=-\gamma$ and $\mathrm{d} \ln \rho(r)/\mathrm{d} \ln r=-2(r/r_{-2})^\alpha$ respectively. Gravitational redshift
profiles can be calculated using these analytic density profiles.

For NFW density profiles, we need to assume an inner cut-off radius $R_t$ to
prevent logarithmic divergence.
\begin{equation}
\label{nfw_z_int}
\begin{split}
\phi(r) &= -4\pi G\left[\frac 1 r \int_0^r\;\rho(r)r^2\mathrm{d}r+\int_r^{R_t}\;\rho(r)r\mathrm{d}r\right]\\
& = \frac {4\pi G\rho_\mathrm{crit}(z)\delta_c r^3_s} {r_s+R_t}\left[\frac {r_s+R_t} r\ln\left(\frac {r_s} {r+r_s}\right)+1\right]\\
& = \frac {4\pi G\rho_\mathrm{crit}(z)\delta_c r^3_s} r \ln\left(\frac {r_s} {r+r_s}\right)+\phi_0~.
\end{split}
\end{equation}
Since the second term of line 2 in Eqn.~\ref{nfw_z_int} is a constant,
we can take only the first term and add a constant $\phi_0$ to obtain
line 3. As expected, the $r$-dependent term in line 3 is
independent of $R_t$.

For gNFW density profiles,
\begin{align}\label{gnfw_z_int}
\nonumber \phi(r) = & \frac {4\pi G\rho_\mathrm{crit}(z) \delta_c } {r^{\gamma-2}} \bigg[\frac{r_s^\gamma{}_2F_1\left(3-\gamma,3-\gamma;4-\gamma;-r/ {r_s}\right)} {\gamma-3} \\
&+\frac {r_s^2\left(r+r_s\right)^{\gamma-2}} {2-\gamma}\bigg]+\phi'_0,
\end{align}
where ${}_2F_1\left(a,b;c;z\right)$ is the hypergeometric function. 
The NFW density profile is therefore a special case of the
gNFW density profile when $\gamma=1$.

For Einasto density profiles, there is no analytic solution for the potential,
and so we carry out the required integration numerically.

In order to apply these results to the gravitational potential profiles
of halos in our simulations, 
we start by fitting the density profiles of the massive halos. 
Following \cite{neto2007}, we minimize the mean
squared deviation between the binned data profile $\rho$ and the model
density profile $\rho_\mathrm{model}(\Theta)$ with a set of fitting
parameters $\Theta$ in log-log space, defined as
\begin{equation} \label{eqn:fit}
\sigma_\mathrm{density}^2 = \frac 1 {N_b - 1} \sum\limits_{i=1}^{N_b}\left(\log_{10}\rho_i - \log_{10}\rho_{\mathrm{model},i}(\Theta)\right)^2,
\end{equation}
where $N_b$ is the number of radial bins. Eqn.~\ref{eqn:fit} is valid
only when each bin has the same weight. We have also tried weighting each bin 
by
the inverse variance due to Poisson noise  but there is significant
difference in our results. The parameter
sets are $\Theta=(\delta_c, r_s)$ for NFW density profiles,
$\Theta=(\delta_c, r_s, \gamma)$ for gNFW density profiles and
$\Theta=(\alpha, r_{-2}, \rho_{-2})$ for Einasto density
profiles. After we fit the density profiles using the different models,
we use the output parameters to calculate the  $z_\mathrm{g}$ profiles. The
evaluation of the goodness of fit for the $z_\mathrm{g}$ profiles is given by
\begin{equation}\label{eqn:grav_fit}
\sigma_\mathrm{grav}^2 = \frac 1 {N_b - 1} \sum\limits_{i=1}^{N_b}\left(z_{\mathrm{g},i} - z_{\mathrm{g},\mathrm{model},i}(\Theta)\right)^2~.
\end{equation}

Dark matter halos, particularly of
large mass are dynamic structures. Even though we
might expect to see more relaxed halos at low redshift,
Fig.~\ref{fig:sliceplot} contains obvious substructure. Looking at 
all halos, irrespective of mass, there are  cases among them where 
ongoing merging is obvious. 
It is therefore useful to quantify which halos are close to equilibrium
 states. Following \citet{thomas2001} and \citet{neto2007}, we measure the
degree of substructure based on the center of mass displacement,
defining
\begin{equation} \label{eqn:norm_off}
s = \frac {|\bm{r}_c-\bm{r}_\mathrm{cm}|} {R_\mathrm{vir}},
\end{equation}
where $\bm{r}_c$ denotes the potential minimum and
$\bm{r}_\mathrm{cm}$ denotes the center of mass. $s$ is referred to as the
normalized offset. A \textit{relaxed} halo requires $s<0.07$ to
meet that definition. We have
computed $s$ for each of the ten most massive halos in 
the simulation and find one
merger. We have eliminated it from Tbl.~2.

\begin{figure*}
  \centering
    \includegraphics[width=1\textwidth]{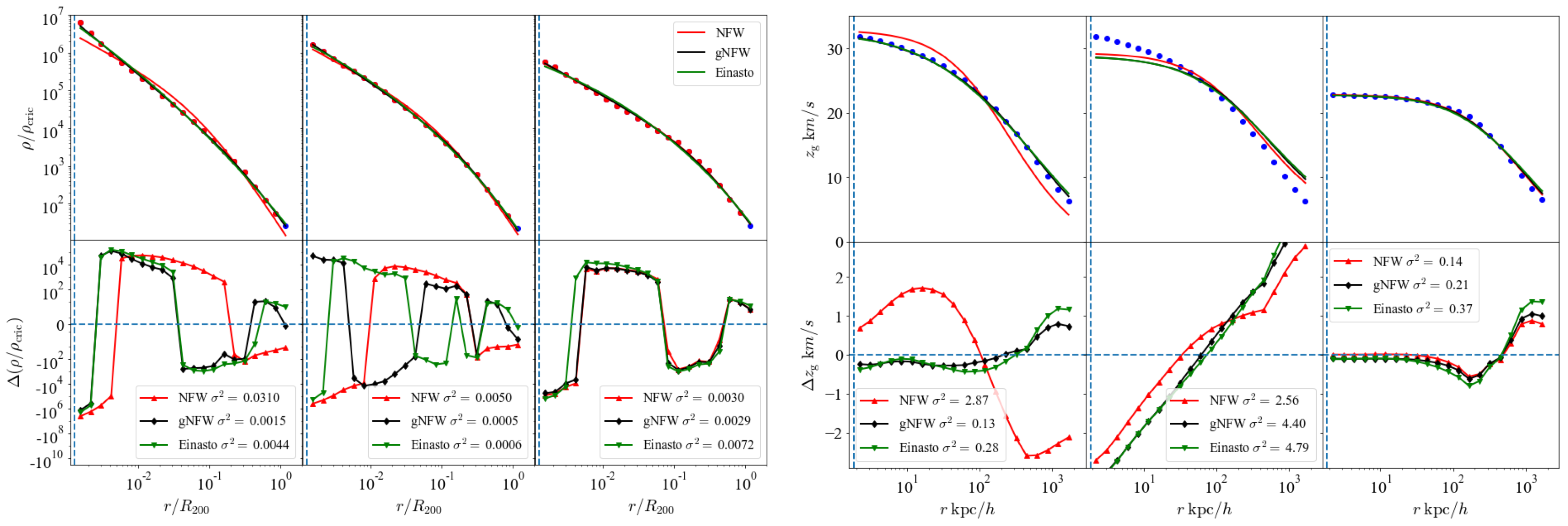}
    \caption{Halo density and gravitational redshift
profiles for different types of particle and 
their analytic fits. The set of figures contains two blocks, from left to right,
 and each block
      contains six subplots. The left and right blocks
      illustrate fits to the density profiles and calculated gravitational redshift profiles measured from the particle
data respectively. Inside each block, the top panels show fits to  
all types of particles in MBII (top left), dark
      matter particles in MBII (top middle), DMO (top right) and bottom panels show the differences between data and corresponding models.
      All plots are for the halo which in the MBII simulations has halo mass $
      7\times10^{14}\;M_\odot/h$. The center is chosen at the
      potential minimum in the halo. In the top right panel, the blue
      filled circles show the density profile of the halo. After
taking into consideration 
      the softening length, $1.85\,\mathrm{kpc}/h$, data between
      $2\,\mathrm{kpc}/h$ and $r_{200}$ (the red filled circles) are
      used in the fit.}
     \label{fig:fitting}
\end{figure*}

Fig.~\ref{fig:fitting} shows density and gravitational redshift profiles of the
most massive halo in the MBII catalogue as well as the corresponding
mean squared errors on the fit ($\sigma^2$s). The two blocks of figures
 (each block
has six panels) show fits to all types of particles in MBII, dark
matter particles in MBII and DMO respectively. Looking at the left
block first, we can see that the density fits from all three blocks 
(especially the
bottom left panels in each block) demonstrate a consistent trend:
the NFW profile is good in the outer regions but underestimates the
density in the innermost regions. Not surprisingly, the NFW profile fits
the DMO simulation
the best, but both gNFW and Einasto density profiles do a better job as
they have one more fit parameter. The right block also provides us with
information on the influence of baryons. It is interesting to
see that fitting parameters from the NFW relation applied
to all types of particles
in MBII (left panels on the right block) overestimate $z_g$ near the center and
underestimate $z_g$ away from the center while they behave in quite the
opposite fashion when applied to  dark matter particles in 
MBII (middle panels on the right block). Among the three
models, gNFW yields the best performance. We therefore propose that the
most useful analytic gravitational redshift profile is the one 
based on the gNFW density profile, given by
Eqn.~\ref{gnfw_z_int}.

%

\begin{figure}
  \centering
    \includegraphics[width=1\columnwidth]{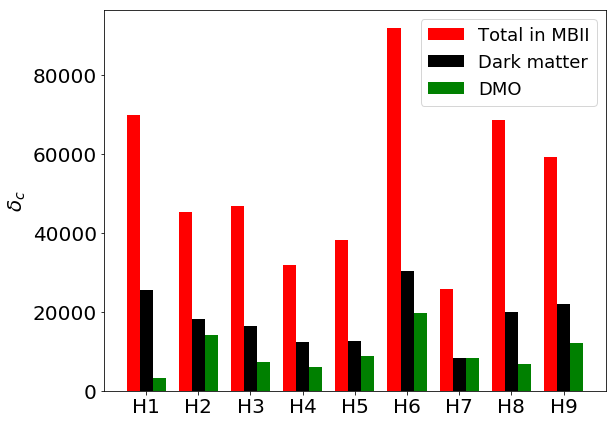}
    \caption{A bar plot showing the concentration $\delta_c$  for
the 9 most massive dynamically relaxed halos in the simulation. We give 
results for all
      particles (red), dark matter (black) in MBII and in dark
      matter-only simulations (green). Labels along $x$-axis indicate
      the different halos.}
     \label{fig:barplot_params}
\end{figure}

So that the reader can
reproduce the fit curves exactly, the parameters are listed in 
Tbl.~2 of the Appendix. We plot the concentration parameter $\delta_c$
for all 9 halos in our study in Fig.~\ref{fig:barplot_params}.
We can see that in each block,
the most concentrated distribution is that of all types of particles
in MBII while the least concentrated one is that of particles in
DMO. This again shows that the effect of baryons is important in the
inner regions, with the concentration increasing by almost an order
of magnitude in some systems.

\section{Velocities and the transverse Doppler effect}
\label{sec:velocity}
In virialized objects such as dark matter halos, 
the gravitational redshift is supplemented by an additional
component of comparable magnitude (\citealt{zhao2013}).
The observed wavelength of photons at
redshift $z$ is affected by the following relation,

\begin{equation}
\frac{c}{1+z}\frac{\lambda_\mathrm{obs}}{\lambda_\mathrm{emit}}=\left[c+\frac{{\phi}-\bm{v}^2/2}{c}\right],
\label{eq:td}
\end{equation}
where $\phi$ is the potential and $\bm{v}$ is the peculiar velocity of the
location from which the photon is emitted. The additional velocity
term is known as the Transverse Doppler (TD) effect, one of the main novel
predictions of the Special Theory of Relativity related to  
object motion.

\begin{figure}
  \centering
    \includegraphics[width=1\columnwidth]{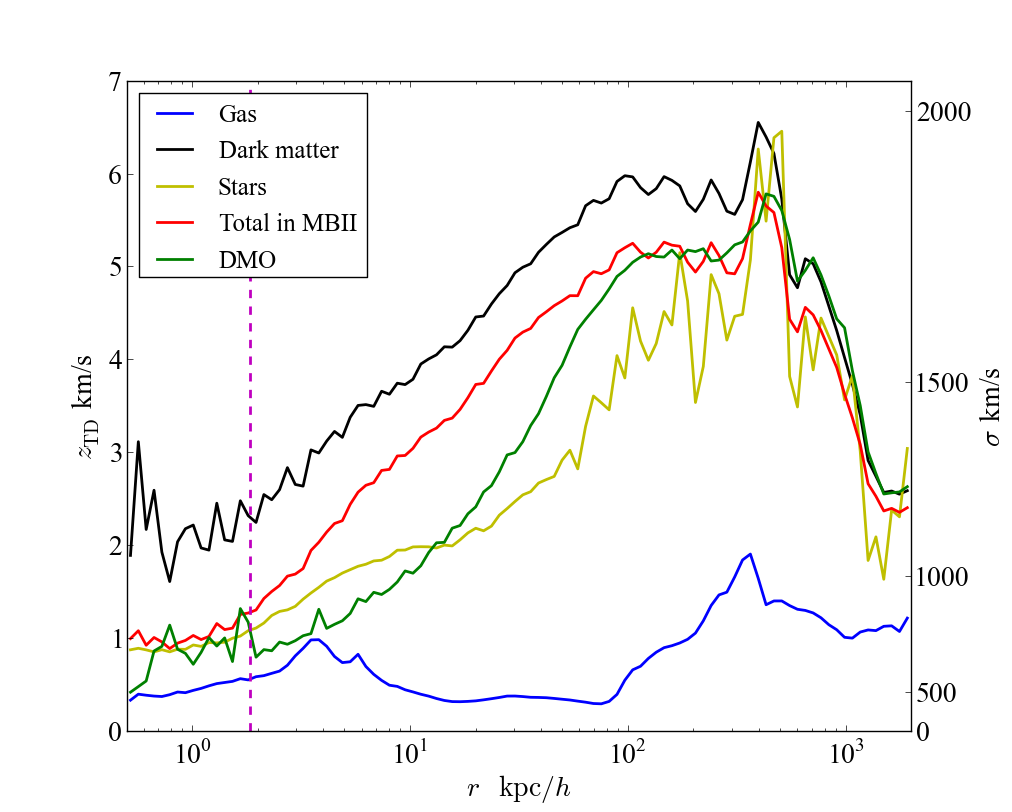}
    \caption{The transverse Doppler redshift profiles of gas (blue), dark
      matter (black), stars (yellow) and their average weighted by mass
      (red) in MBII and dark matter (green) in the DMO simulations. Results
are for the  
      halo of mass $7\times10^{14}\;M_\odot/h$. The center is chosen
      at the potential minimum of the halo. The distance $r$ spans
      slightly above the gravitational softening $\epsilon =
      1.85\;\mathrm{kpc}/h$ to $2\;\mathrm{Mpc}/h$.}
     \label{fig:td}
\end{figure}

We have computed the TD effect profile in a similar
fashion to the gravitational redshift profile, averaging
the mass-weighted TD redshift in spherical cells centered on the center
of each halo (chosen to be the minimum of the gravitational potential).
Fig.~\ref{fig:td} shows the TD  profile of
different components for the most massive halo. 
In general, the TD effect leads to a positive addition to the redshift, 
but with the outskirts being more redshifted than the halo center.
This is opposite to the
gravitational redshift, and therefore acts to suppress the overall
redshift signal. The TD amplitude (difference between center and
virial radius) is about 4 km/s, which is
about 25\% of the gravitational redshift. In detail, the TD effect 
increases with radius
below 100 kpc/$h$ and decreases with radius above 100 kpc/$h$. 

This trend can be interpreted by assuming an isotropic power-law density
distribution $\rho\propto r^{-\alpha}$.  The enclosed mass is then
$M(r)=\int\mathrm{d}r4\pi r^2\rho\propto r^{3-\alpha}$. The
tangential  component of the velocity is therefore
 $v_t=\sqrt{GM(r)/r}\propto r^{1-\alpha/2}$, which leads to a TD
 effect profile $\bm{v}^2/2=(3v_t)^2\propto
r^{2-\alpha}$. This  relation shows the TD effect
is sensitive to the slope of the density profile. Referring to the
blue curve in the left panel of Fig.~\ref{fig:projected}, $\alpha>2$
when $r$ is large, while $\alpha$ becomes small when it comes closer to
the core. This therefore qualitatively explains  the 
transverse Doppler velocity
profile shown in Fig.~\ref{fig:td}.

The  different lines in Fig.~\ref{fig:td} show the TD profiles measured
from the different components in the MBII simulation, as well as the 
profile from the DMO simulation. In MBII, all components feel the same
gravitational potential, and so one might expect the velocity dispersions,
and hence the magnitude of the TD effect to be the same for all components.
In fact, we see that the TD effect for the stars is approximately $50$\%
smaller than for the dark matter in the central regions of the halo, and 
$\sim 25$\% lower around the virial radius. 
The red curve is the mass weighted average of all 
components and there is a $\sim 10-15$\% difference in TD between this
and stars at the virial radius. This difference is related to the ``velocity
bias'' that has been seen between galaxies and dark matter as tracers
of the large-scale density field. \cite{Armitage2018}
have found in galaxy clusters that mass estimates made from the
velocity dispersion of stellar mass selected
galaxies are only minimally affected by velocity bias (i.e., similar to
estimates made using the dark matter velocities to within 5\%). On the
other hand, \cite{Ye2017} have shown how central
galaxies in halos have a small (but still non negligible)
velocity dispersion with respect to the dark
matter. 

In our case, because of the small numbers of subhalos in each halo,  
we have not computed the TD profiles after separating
 the mass in the different components
into subhalos or galaxies. The possibility of
 differences in the TD effect between 
stars and dark matter should be borne in mind when comparing to
observations, as should the effects of baryons overall.  We do
find, however that the effects of baryons on the TD effect are smaller
that for the gravitational redshift. 
We can see this by comparing the stars curve (yellow) to the DMO profile 
(green). We  see that on scales between
$\sim$ 2-10 $\mathrm{kpc}$ the stars are boosted by about 0.5 km/s in
$z_\mathrm{TD}$, which corresponds to  a $\sim0.5$  km/s decrement in the overall
redshift profile $z_g$ (as
$z_\mathrm{TD}$ has the opposite sign from Eqn.~\ref{eq:td}). On larger scales,
up to the virial radius, the sign changes, but the effect is of similar 
magnitude.

\section{Conclusions}\label{sec:discussion}

We have explored the effects of baryons on the density,
gravitational redshift and velocity profiles of the 9 most massive
$\Lambda$CDM halos in the MBII hydrodynamical and dark matter-only (DMO)
simulations. In the hydro run (MBII) we have studied the differences
between dark matter profiles and all-particle profiles. We have also
cross-compared dark matter profiles and all-particle profiles in MBII
with the profiles from DMO. On large scales (several tens of kpc/$h$) we
find good agreement among all three profiles (black, red and green
lines in Fig.~\ref{fig:density}, Fig.~\ref{fig:z_types} and
Fig.~\ref{fig:td}, indicating that a reasonably good approximation to the
profiles can be obtained using
 $N$-body simulations, as has been done in
\cite{zhao2013, cai2016, Zhu2016Nbody}.

On small scales we  observe deviations from the $N$-body only simulations
due to the clustering of baryons (gas and stellar particles).
The baryons have excess concentration
 toward the center
bringing  dark matter particles inward, making the inner profile
steeper than $N$-body simulations. This  leads to a significant
difference, an extra $\sim 8$ km/s in gravitational redshift compared to the
total ~16 km/s difference between center and outskirt in $N$-body
simulations of cluster sized halos.
We  fit density profiles with NFW, gNFW and Einasto
relations to find quantitatively the differences for simulations
 with and without
baryons. We find that baryons  make density profiles 30\%-50\% more
concentrated ($c_{200}$).

We have examined  the differences between projected
redshift profiles (which can be measured more easily from observations)
 and spatial profiles. We
find that projected profiles, by including information from the
edges of systems tend to resemble spatial profiles at larger radius.
We have also studied the transverse Doppler effect which is expected to
be a
comparable effect in magnitude to the gravitational redshift. 
We find that baryonic effects are relatively small, at the $\sim 0.5$ km/s 
level. 

Overall, when considering the gravitational redshift profiles
of galaxy clusters, such as those presented in \cite{wojtak2011,zhao2013,bonvin2014,Zhu2016Nbody,Alam2016Measurement,Giusarma2016PT,cai2016},
one should consider the possiblity of baryonic and galaxy formation
effects. These physical processes are those also relevant 
to mass determinations of galaxy
clusters \citep{neto2007, Rudd2008,duffy2010,Gnedin2011,Fedeli2012,Velliscig2014,chan2015,schaller2015, Schaller2015b}

We have learnt in the present work that  baryonic
effects are most important on small scales,  in
the centers of clusters. However, there are  possible extensions to this
work. For example, there are several other relativistic effects of the
same order of magnitude which can affect the overall redshift 
profiles of clusters. These include the light cone effect and relativistic
beaming effect mentioned by \cite{kaiser2013}, \cite{Zhu2016Nbody},
\cite{Alam2016TS} and \cite{Breton2018}. These effects are all
potentially affected by the presence of baryons. In order to 
properly match the
observations, one must consider a full model in which all these
effects are taken into account. In addition, the mismatch between the
hydrodynamic MBII simulation and the $N$-body only DMO
run happens on galaxy scales, and on the scales of galaxy potentials.
As measurements of gravitational redshifts are made using 
 the differences between galaxy redshifts, simulations that explore
the structure in galaxy potentials at higher resolution are also needed.

\section*{Acknowledgments}
This work was supported by NSF Award AST-1412966. SA is also
supported by the European Research Council through the COSFORM
Research Grant (\#670193). RACC would like to thank the Astrophysics group at the University of Melbourne for their hospitality and the support of a Lyle Fellowship. We thank Nishikanta Khandai and Ananth Tenneti for providing halo and subhalo catalogues of the MassiveBlack II and MassiveBlack II dark matter-only simulations.


\bibliography{Master_Hongyu.bib}

\begin{thebibliography}{}
\makeatletter
\relax
\def\mn@urlcharsother{\let\do\@makeother \do\$\do\&\do\#\do\^\do\_\do\%\do\~}
\def\mn@doi{\begingroup\mn@urlcharsother \@ifnextchar [ {\mn@doi@}
  {\mn@doi@[]}}
\def\mn@doi@[#1]#2{\def\@tempa{#1}\ifx\@tempa\@empty \href
  {http://dx.doi.org/#2} {doi:#2}\else \href {http://dx.doi.org/#2} {#1}\fi
  \endgroup}
\def\mn@eprint#1#2{\mn@eprint@#1:#2::\@nil}
\def\mn@eprint@arXiv#1{\href {http://arxiv.org/abs/#1} {{\tt arXiv:#1}}}
\def\mn@eprint@dblp#1{\href {http://dblp.uni-trier.de/rec/bibtex/#1.xml}
  {dblp:#1}}
\def\mn@eprint@#1:#2:#3:#4\@nil{\def\@tempa {#1}\def\@tempb {#2}\def\@tempc
  {#3}\ifx \@tempc \@empty \let \@tempc \@tempb \let \@tempb \@tempa \fi \ifx
  \@tempb \@empty \def\@tempb {arXiv}\fi \@ifundefined
  {mn@eprint@\@tempb}{\@tempb:\@tempc}{\expandafter \expandafter \csname
  mn@eprint@\@tempb\endcsname \expandafter{\@tempc}}}

\bibitem[\protect\citeauthoryear{Abel}{Abel}{1826}]{abel1826auflosung}
Abel N.,  1826, Journal f{\"u}r die reine und angewandte Mathematik, 1, 153

\bibitem[\protect\citeauthoryear{{Alam}, {Zhu}, {Croft}, {Ho}, {Giusarma}  \&
  {Schneider}}{{Alam} et~al.}{2017a}]{Alam2016Measurement}
{Alam} S.,  {Zhu} H.,  {Croft} R.~A.~C.,  {Ho} S.,  {Giusarma} E.,
  {Schneider} D.~P.,  2017a, \mn@doi [\mnras] {10.1093/mnras/stx1421}, \href
  {http://adsabs.harvard.edu/abs/2017MNRAS.470.2822A} {470, 2822}

\bibitem[\protect\citeauthoryear{{Alam}, {Croft}, {Ho}, {Zhu}  \&
  {Giusarma}}{{Alam} et~al.}{2017b}]{Alam2016TS}
{Alam} S.,  {Croft} R.~A.~C.,  {Ho} S.,  {Zhu} H.,   {Giusarma} E.,  2017b,
  \mn@doi [\mnras] {10.1093/mnras/stx1684}, \href
  {http://adsabs.harvard.edu/abs/2017MNRAS.471.2077A} {471, 2077}

\bibitem[\protect\citeauthoryear{{Armitage}, {Barnes}, {Kay}, {Bah{\'e}},
  {Dalla Vecchia}, {Crain}  \& {Theuns}}{{Armitage}
  et~al.}{2018}]{Armitage2018}
{Armitage} T.~J.,  {Barnes} D.~J.,  {Kay} S.~T.,  {Bah{\'e}} Y.~M.,  {Dalla
  Vecchia} C.,  {Crain} R.~A.,   {Theuns} T.,  2018, \mn@doi [\mnras]
  {10.1093/mnras/stx3020}, \href
  {http://adsabs.harvard.edu/abs/2018MNRAS.474.3746A} {474, 3746}

\bibitem[\protect\citeauthoryear{{Bonvin}, {Hui}  \& {Gazta{\~n}aga}}{{Bonvin}
  et~al.}{2014}]{bonvin2014}
{Bonvin} C.,  {Hui} L.,   {Gazta{\~n}aga} E.,  2014, \mn@doi [\prd]
  {10.1103/PhysRevD.89.083535}, \href
  {http://adsabs.harvard.edu/abs/2014PhRvD..89h3535B} {89, 083535}

\bibitem[\protect\citeauthoryear{{Breton}, {Rasera}, {Taruya}, {Lacombe}  \&
  {Saga}}{{Breton} et~al.}{2018}]{Breton2018}
{Breton} M.-A.,  {Rasera} Y.,  {Taruya} A.,  {Lacombe} O.,   {Saga} S.,  2018,
  preprint, \href {http://adsabs.harvard.edu/abs/2018arXiv180304294B} {}
  (\mn@eprint {arXiv} {1803.04294})

\bibitem[\protect\citeauthoryear{{Cai}, {Kaiser}, {Cole}  \& {Frenk}}{{Cai}
  et~al.}{2017}]{cai2016}
{Cai} Y.-C.,  {Kaiser} N.,  {Cole} S.,   {Frenk} C.,  2017, \mn@doi [\mnras]
  {10.1093/mnras/stx469}, \href
  {http://adsabs.harvard.edu/abs/2017MNRAS.468.1981C} {468, 1981}

\bibitem[\protect\citeauthoryear{{Cappi}}{{Cappi}}{1995}]{cappi1995}
{Cappi} A.,  1995, \aap, \href
  {http://adsabs.harvard.edu/abs/1995A%26A...301....6C} {301, 6}

\bibitem[\protect\citeauthoryear{{Chan}, {Kere{\v s}}, {O{\~n}orbe}, {Hopkins},
  {Muratov}, {Faucher-Gigu{\`e}re}  \& {Quataert}}{{Chan}
  et~al.}{2015}]{chan2015}
{Chan} T.~K.,  {Kere{\v s}} D.,  {O{\~n}orbe} J.,  {Hopkins} P.~F.,  {Muratov}
  A.~L.,  {Faucher-Gigu{\`e}re} C.-A.,   {Quataert} E.,  2015, \mn@doi [\mnras]
  {10.1093/mnras/stv2165}, \href
  {http://adsabs.harvard.edu/abs/2015MNRAS.454.2981C} {454, 2981}

\bibitem[\protect\citeauthoryear{{Croft}}{{Croft}}{2013}]{croft2013}
{Croft} R.~A.~C.,  2013, \mn@doi [\mnras] {10.1093/mnras/stt1223}, \href
  {http://adsabs.harvard.edu/abs/2013MNRAS.434.3008C} {434, 3008}

\bibitem[\protect\citeauthoryear{{Davis}, {Efstathiou}, {Frenk}  \&
  {White}}{{Davis} et~al.}{1985}]{davis1985}
{Davis} M.,  {Efstathiou} G.,  {Frenk} C.~S.,   {White} S.~D.~M.,  1985,
  \mn@doi [\apj] {10.1086/163168}, \href
  {http://adsabs.harvard.edu/abs/1985ApJ...292..371D} {292, 371}

\bibitem[\protect\citeauthoryear{Dodelson}{Dodelson}{2003}]{dodelson2003modern}
Dodelson S.,  2003, Modern cosmology.
Academic press

\bibitem[\protect\citeauthoryear{{Duffy}, {Schaye}, {Kay}, {Dalla Vecchia},
  {Battye}  \& {Booth}}{{Duffy} et~al.}{2010}]{duffy2010}
{Duffy} A.~R.,  {Schaye} J.,  {Kay} S.~T.,  {Dalla Vecchia} C.,  {Battye}
  R.~A.,   {Booth} C.~M.,  2010, \mn@doi [\mnras]
  {10.1111/j.1365-2966.2010.16613.x}, \href
  {http://adsabs.harvard.edu/abs/2010MNRAS.405.2161D} {405, 2161}

\bibitem[\protect\citeauthoryear{{Einasto}}{{Einasto}}{1965}]{einasto1965}
{Einasto} J.,  1965, Trudy Astrofizicheskogo Instituta Alma-Ata, \href
  {http://adsabs.harvard.edu/abs/1965TrAlm...5...87E} {5, 87}

\bibitem[\protect\citeauthoryear{{Einstein}}{{Einstein}}{1916}]{Einstein1916}
{Einstein} A.,  1916, Annalen Der Physik, \href
  {http://adsabs.harvard.edu/cgi-bin/nph-data_query?bibcode=1916AnP...354..769E&link_type=ARTICLE&db_key=AST&high=}
  {49, 770}

\bibitem[\protect\citeauthoryear{{Falcon}, {Winget}, {Montgomery}  \&
  {Williams}}{{Falcon} et~al.}{2010}]{falcon2010}
{Falcon} R.~E.,  {Winget} D.~E.,  {Montgomery} M.~H.,   {Williams} K.~A.,
  2010, \mn@doi [\apj] {10.1088/0004-637X/712/1/585}, \href
  {http://adsabs.harvard.edu/abs/2010ApJ...712..585F} {712, 585}

\bibitem[\protect\citeauthoryear{{Fedeli}}{{Fedeli}}{2012}]{Fedeli2012}
{Fedeli} C.,  2012, \mn@doi [\mnras] {10.1111/j.1365-2966.2012.21302.x}, \href
  {http://adsabs.harvard.edu/abs/2012MNRAS.424.1244F} {424, 1244}

\bibitem[\protect\citeauthoryear{{Giusarma}, {Alam}, {Zhu}, {Croft}  \&
  {Ho}}{{Giusarma} et~al.}{2016}]{Giusarma2016PT}
{Giusarma} E.,  {Alam} S.,  {Zhu} H.,  {Croft} R.~A.~C.,   {Ho} S.,  2016,
  \mn@doi [xxx] {xxxx}, \href {xx} {pp xx--xx}

\bibitem[\protect\citeauthoryear{{Gnedin}, {Ceverino}, {Gnedin}, {Klypin},
  {Kravtsov}, {Levine}, {Nagai}  \& {Yepes}}{{Gnedin}
  et~al.}{2011}]{Gnedin2011}
{Gnedin} O.~Y.,  {Ceverino} D.,  {Gnedin} N.~Y.,  {Klypin} A.~A.,  {Kravtsov}
  A.~V.,  {Levine} R.,  {Nagai} D.,   {Yepes} G.,  2011, preprint, \href
  {http://adsabs.harvard.edu/abs/2011arXiv1108.5736G} {} (\mn@eprint {arXiv}
  {1108.5736})

\bibitem[\protect\citeauthoryear{{Hernquist}}{{Hernquist}}{1990}]{hernquist1990}
{Hernquist} L.,  1990, \mn@doi [\apj] {10.1086/168845}, \href
  {http://adsabs.harvard.edu/abs/1990ApJ...356..359H} {356, 359}

\bibitem[\protect\citeauthoryear{{Huang}, {Eifler}, {Mandelbaum}  \&
  {Dodelson}}{{Huang} et~al.}{2018}]{Huang2018}
{Huang} H.-J.,  {Eifler} T.,  {Mandelbaum} R.,   {Dodelson} S.,  2018,
  preprint, \href {http://adsabs.harvard.edu/abs/2018arXiv180901146H} {}
  (\mn@eprint {arXiv} {1809.01146})

\bibitem[\protect\citeauthoryear{{Kaiser}}{{Kaiser}}{2013}]{kaiser2013}
{Kaiser} N.,  2013, \mn@doi [\mnras] {10.1093/mnras/stt1370}, \href
  {http://adsabs.harvard.edu/abs/2013MNRAS.435.1278K} {435, 1278}

\bibitem[\protect\citeauthoryear{{Khandai}, {Di Matteo}, {Croft}, {Wilkins},
  {Feng}, {Tucker}, {DeGraf}  \& {Liu}}{{Khandai} et~al.}{2015}]{khandai2015}
{Khandai} N.,  {Di Matteo} T.,  {Croft} R.,  {Wilkins} S.,  {Feng} Y.,
  {Tucker} E.,  {DeGraf} C.,   {Liu} M.-S.,  2015, \mn@doi [\mnras]
  {10.1093/mnras/stv627}, \href
  {http://adsabs.harvard.edu/abs/2015MNRAS.450.1349K} {450, 1349}

\bibitem[\protect\citeauthoryear{{Kim} \& {Croft}}{{Kim} \&
  {Croft}}{2004}]{kim2004}
{Kim} Y.-R.,  {Croft} R.~A.~C.,  2004, \mn@doi [\apj] {10.1086/383218}, \href
  {http://adsabs.harvard.edu/abs/2004ApJ...607..164K} {607, 164}

\bibitem[\protect\citeauthoryear{{McDonald}}{{McDonald}}{2009}]{mcDonald2009}
{McDonald} P.,  2009, \mn@doi [\jcap] {10.1088/1475-7516/2009/11/026}, \href
  {http://adsabs.harvard.edu/abs/2009JCAP...11..026M} {11, 026}

\bibitem[\protect\citeauthoryear{{Merritt}, {Navarro}, {Ludlow}  \&
  {Jenkins}}{{Merritt} et~al.}{2005}]{merritt2005}
{Merritt} D.,  {Navarro} J.~F.,  {Ludlow} A.,   {Jenkins} A.,  2005, \mn@doi
  [\apjl] {10.1086/430636}, \href
  {http://adsabs.harvard.edu/abs/2005ApJ...624L..85M} {624, L85}

\bibitem[\protect\citeauthoryear{{Merritt}, {Graham}, {Moore}, {Diemand}  \&
  {Terzi{\'c}}}{{Merritt} et~al.}{2006}]{merritt2006}
{Merritt} D.,  {Graham} A.~W.,  {Moore} B.,  {Diemand} J.,   {Terzi{\'c}} B.,
  2006, \mn@doi [\aj] {10.1086/508988}, \href
  {http://adsabs.harvard.edu/abs/2006AJ....132.2685M} {132, 2685}

\bibitem[\protect\citeauthoryear{{Moore}, {Quinn}, {Governato}, {Stadel}  \&
  {Lake}}{{Moore} et~al.}{1999}]{moore1999}
{Moore} B.,  {Quinn} T.,  {Governato} F.,  {Stadel} J.,   {Lake} G.,  1999,
  \mn@doi [\mnras] {10.1046/j.1365-8711.1999.03039.x}, \href
  {http://adsabs.harvard.edu/abs/1999MNRAS.310.1147M} {310, 1147}

\bibitem[\protect\citeauthoryear{{Navarro}, {Frenk}  \& {White}}{{Navarro}
  et~al.}{1996}]{navarro1996}
{Navarro} J.~F.,  {Frenk} C.~S.,   {White} S.~D.~M.,  1996, \mn@doi [\apj]
  {10.1086/177173}, \href {http://adsabs.harvard.edu/abs/1996ApJ...462..563N}
  {462, 563}

\bibitem[\protect\citeauthoryear{{Navarro}, {Frenk}  \& {White}}{{Navarro}
  et~al.}{1997}]{navarro1997}
{Navarro} J.~F.,  {Frenk} C.~S.,   {White} S.~D.~M.,  1997, \apj, \href
  {http://adsabs.harvard.edu/abs/1997ApJ...490..493N} {490, 493}

\bibitem[\protect\citeauthoryear{{Navarro} et~al.,}{{Navarro}
  et~al.}{2004}]{navarro2004}
{Navarro} J.~F.,  et~al., 2004, \mn@doi [\mnras]
  {10.1111/j.1365-2966.2004.07586.x}, \href
  {http://adsabs.harvard.edu/abs/2004MNRAS.349.1039N} {349, 1039}

\bibitem[\protect\citeauthoryear{{Neto} et~al.,}{{Neto}
  et~al.}{2007}]{neto2007}
{Neto} A.~F.,  et~al., 2007, \mn@doi [\mnras]
  {10.1111/j.1365-2966.2007.12381.x}, \href
  {http://adsabs.harvard.edu/abs/2007MNRAS.381.1450N} {381, 1450}

\bibitem[\protect\citeauthoryear{{Pound} \& {Rebka}}{{Pound} \&
  {Rebka}}{1959}]{pound1959}
{Pound} R.~V.,  {Rebka} G.~A.,  1959, \mn@doi [Physical Review Letters]
  {10.1103/PhysRevLett.3.439}, \href
  {http://adsabs.harvard.edu/abs/1959PhRvL...3..439P} {3, 439}

\bibitem[\protect\citeauthoryear{{Rudd}, {Zentner}  \& {Kravtsov}}{{Rudd}
  et~al.}{2008}]{Rudd2008}
{Rudd} D.~H.,  {Zentner} A.~R.,   {Kravtsov} A.~V.,  2008, \mn@doi [\apj]
  {10.1086/523836}, \href {http://adsabs.harvard.edu/abs/2008ApJ...672...19R}
  {672, 19}

\bibitem[\protect\citeauthoryear{{Schaller} et~al.,}{{Schaller}
  et~al.}{2015a}]{schaller2015}
{Schaller} M.,  et~al., 2015a, \mn@doi [\mnras] {10.1093/mnras/stv1067}, \href
  {http://adsabs.harvard.edu/abs/2015MNRAS.451.1247S} {451, 1247}

\bibitem[\protect\citeauthoryear{{Schaller} et~al.,}{{Schaller}
  et~al.}{2015b}]{Schaller2015b}
{Schaller} M.,  et~al., 2015b, \mn@doi [\mnras] {10.1093/mnras/stv1341}, \href
  {http://adsabs.harvard.edu/abs/2015MNRAS.452..343S} {452, 343}

\bibitem[\protect\citeauthoryear{{Schmidt} \& {Allen}}{{Schmidt} \&
  {Allen}}{2007}]{schmidt2007}
{Schmidt} R.~W.,  {Allen} S.~W.,  2007, \mn@doi [\mnras]
  {10.1111/j.1365-2966.2007.11928.x}, \href
  {http://adsabs.harvard.edu/abs/2007MNRAS.379..209S} {379, 209}

\bibitem[\protect\citeauthoryear{{S{\'e}rsic}}{{S{\'e}rsic}}{1963}]{sersic1963}
{S{\'e}rsic} J.~L.,  1963, Boletin de la Asociacion Argentina de Astronomia La
  Plata Argentina, \href {http://adsabs.harvard.edu/abs/1963BAAA....6...41S}
  {6, 41}

\bibitem[\protect\citeauthoryear{{Springel}}{{Springel}}{2005}]{springel2005}
{Springel} V.,  2005, \mn@doi [\mnras] {10.1111/j.1365-2966.2005.09655.x},
  \href {http://adsabs.harvard.edu/abs/2005MNRAS.364.1105S} {364, 1105}

\bibitem[\protect\citeauthoryear{{Springel}, {Yoshida}  \& {White}}{{Springel}
  et~al.}{2001}]{springel2001}
{Springel} V.,  {Yoshida} N.,   {White} S.~D.~M.,  2001, \mn@doi [\na]
  {10.1016/S1384-1076(01)00042-2}, \href
  {http://adsabs.harvard.edu/abs/2001NewA....6...79S} {6, 79}

\bibitem[\protect\citeauthoryear{{Takeda} \& {Ueno}}{{Takeda} \&
  {Ueno}}{2012}]{takeda2012}
{Takeda} Y.,  {Ueno} S.,  2012, \mn@doi [\solphys] {10.1007/s11207-012-0068-8},
  \href {http://adsabs.harvard.edu/abs/2012SoPh..281..551T} {281, 551}

\bibitem[\protect\citeauthoryear{{Tenneti}, {Mandelbaum}, {Di Matteo},
  {Kiessling}  \& {Khandai}}{{Tenneti} et~al.}{2015}]{tenneti2015}
{Tenneti} A.,  {Mandelbaum} R.,  {Di Matteo} T.,  {Kiessling} A.,   {Khandai}
  N.,  2015, \mn@doi [\mnras] {10.1093/mnras/stv1625}, \href
  {http://adsabs.harvard.edu/abs/2015MNRAS.453..469T} {453, 469}

\bibitem[\protect\citeauthoryear{{Thomas}, {Muanwong}, {Pearce}, {Couchman},
  {Edge}, {Jenkins}  \& {Onuora}}{{Thomas} et~al.}{2001}]{thomas2001}
{Thomas} P.~A.,  {Muanwong} O.,  {Pearce} F.~R.,  {Couchman} H.~M.~P.,  {Edge}
  A.~C.,  {Jenkins} A.,   {Onuora} L.,  2001, \mn@doi [\mnras]
  {10.1046/j.1365-8711.2001.04330.x}, \href
  {http://adsabs.harvard.edu/abs/2001MNRAS.324..450T} {324, 450}

\bibitem[\protect\citeauthoryear{{Velliscig}, {van Daalen}, {Schaye},
  {McCarthy}, {Cacciato}, {Le Brun}  \& {Dalla Vecchia}}{{Velliscig}
  et~al.}{2014}]{Velliscig2014}
{Velliscig} M.,  {van Daalen} M.~P.,  {Schaye} J.,  {McCarthy} I.~G.,
  {Cacciato} M.,  {Le Brun} A.~M.~C.,   {Dalla Vecchia} C.,  2014, \mn@doi
  [\mnras] {10.1093/mnras/stu1044}, \href
  {http://adsabs.harvard.edu/abs/2014MNRAS.442.2641V} {442, 2641}

\bibitem[\protect\citeauthoryear{{Wojtak}, {Hansen}  \& {Hjorth}}{{Wojtak}
  et~al.}{2011}]{wojtak2011}
{Wojtak} R.,  {Hansen} S.~H.,   {Hjorth} J.,  2011, \mn@doi [\nat]
  {10.1038/nature10445}, \href
  {http://adsabs.harvard.edu/abs/2011Natur.477..567W} {477, 567}

\bibitem[\protect\citeauthoryear{{Ye}, {Guo}, {Zheng}  \& {Zehavi}}{{Ye}
  et~al.}{2017}]{Ye2017}
{Ye} J.-N.,  {Guo} H.,  {Zheng} Z.,   {Zehavi} I.,  2017, \mn@doi [\apj]
  {10.3847/1538-4357/aa70e7}, \href
  {http://adsabs.harvard.edu/abs/2017ApJ...841...45Y} {841, 45}

\bibitem[\protect\citeauthoryear{{Yoo}, {Fitzpatrick}  \& {Zaldarriaga}}{{Yoo}
  et~al.}{2009}]{yoo2009}
{Yoo} J.,  {Fitzpatrick} A.~L.,   {Zaldarriaga} M.,  2009, \mn@doi [\prd]
  {10.1103/PhysRevD.80.083514}, \href
  {http://adsabs.harvard.edu/abs/2009PhRvD..80h3514Y} {80, 083514}

\bibitem[\protect\citeauthoryear{{Yoo}, {Hamaus}, {Seljak}  \&
  {Zaldarriaga}}{{Yoo} et~al.}{2012}]{yoo2012}
{Yoo} J.,  {Hamaus} N.,  {Seljak} U.,   {Zaldarriaga} M.,  2012, \mn@doi [\prd]
  {10.1103/PhysRevD.86.063514}, \href
  {http://adsabs.harvard.edu/abs/2012PhRvD..86f3514Y} {86, 063514}

\bibitem[\protect\citeauthoryear{{Zentner}, {Semboloni}, {Dodelson}, {Eifler},
  {Krause}  \& {Hearin}}{{Zentner} et~al.}{2013}]{Zentner2013}
{Zentner} A.~R.,  {Semboloni} E.,  {Dodelson} S.,  {Eifler} T.,  {Krause} E.,
  {Hearin} A.~P.,  2013, \mn@doi [\prd] {10.1103/PhysRevD.87.043509}, \href
  {http://adsabs.harvard.edu/abs/2013PhRvD..87d3509Z} {87, 043509}

\bibitem[\protect\citeauthoryear{{Zhao}}{{Zhao}}{1996}]{zhao1996}
{Zhao} H.,  1996, \mn@doi [\mnras] {10.1093/mnras/278.2.488}, \href
  {http://adsabs.harvard.edu/abs/1996MNRAS.278..488Z} {278, 488}

\bibitem[\protect\citeauthoryear{{Zhao}, {Peacock}  \& {Li}}{{Zhao}
  et~al.}{2013}]{zhao2013}
{Zhao} H.,  {Peacock} J.~A.,   {Li} B.,  2013, \mn@doi [\prd]
  {10.1103/PhysRevD.88.043013}, \href
  {http://adsabs.harvard.edu/abs/2013PhRvD..88d3013Z} {88, 043013}

\bibitem[\protect\citeauthoryear{{Zhu}, {Alam}, {Croft}, {Ho}  \&
  {Giusarma}}{{Zhu} et~al.}{2017}]{Zhu2016Nbody}
{Zhu} H.,  {Alam} S.,  {Croft} R.~A.~C.,  {Ho} S.,   {Giusarma} E.,  2017,
  \mn@doi [\mnras] {10.1093/mnras/stx1644}, \href
  {http://adsabs.harvard.edu/abs/2017MNRAS.471.2345Z} {471, 2345}

\bibitem[\protect\citeauthoryear{{de Blok}}{{de Blok}}{2010}]{de2010}
{de Blok} W.~J.~G.,  2010, \mn@doi [Advances in Astronomy]
  {10.1155/2010/789293}, \href
  {http://adsabs.harvard.edu/abs/2010AdAst2010E...5D} {2010, 789293}

\bibitem[\protect\citeauthoryear{{de Vaucouleurs}}{{de
  Vaucouleurs}}{1948}]{de1948}
{de Vaucouleurs} G.,  1948, Annales d'Astrophysique, \href
  {http://adsabs.harvard.edu/abs/1948AnAp...11..247D} {11, 247}

\makeatother
\end{thebibliography}
\bibliographystyle{mnras}
\section*{Appendix}

\begin{table*}\label{tbl2}
\caption{Outputs of the fitting parameters: Columns: $M_\mathrm{group}$ is the group  mass. $\delta_\mathrm{c}$ and $r_{s}$ are the outputs from 2-param NFW density profile. $c_{200}$ and $r_{200}$ are calculated from $\delta_\mathrm{c}$ and $r_{s}$. $\alpha$, $r_{-2}$ and $\rho_{-2}/\rho_\mathrm{crit}$ are the outputs from Einasto density profile. Rows: there are 9 blocks representing the top 10 massive halos excluding one merger. The first line of each block shows the (derived) fitting parameters with all types of particles, the second line shows the (derived) fitting parameters from dark matter particles and the third line shows the (derived) fitting parameters from DMO.}
\begin{tabular}{c c c c c c c c}\toprule 
$M_\mathrm{group}$ [$10^{10}\,M_\odot/h$] & $r_{200}$ [$\mathrm{kpc}/h$] & $r_{s}$ [$\mathrm{kpc}/h$] & $c_{200}$ & $\delta_\mathrm{c}$ & $\alpha$ & $r_{-2}$ [$\mathrm{kpc}/h$] &$\rho_{-2}/\rho_\mathrm{crit}$ \\ \hline 
$7.19 \times 10^4$ & 1431.6 & 119.5 & 11.98 & $6.99 \times 10^4$ & 0.0816 & 262.2  & $4.87 \times 10^{-5}$ \\
$7.19 \times 10^4$ & 1431.6 & 180.1 & 7.95  & $2.57 \times 10^4$ & 0.1488 & 264.7  & $4.52 \times 10^{-5}$ \\
$7.47 \times 10^4$ & 1449.2 & 448.3 & 3.23  & $3.32 \times 10^3$ & 0.1764 & 549.9  & $1.11 \times 10^{-5}$ \\ \hline
$4.89 \times 10^4$ & 1258.4 & 125.2 & 10.05 & $4.53 \times 10^4$ & 0.0388 & 1093.5 & $1.17 \times 10^{-6}$ \\
$4.89 \times 10^4$ & 1258.4 & 183.3 & 6.86  & $1.81 \times 10^4$ & 0.1070 & 384.8  & $8.47 \times 10^{-6}$ \\
$5.0 \times 10^4$  & 1267.9 & 204.9 & 6.19  & $1.42 \times 10^4$ & 0.1378 & 340.0  & $1.38 \times 10^{-5}$ \\\hline
$3.66 \times 10^4$ & 1143.1 & 112.3 & 10.18 & $4.68 \times 10^4$ & 0.1151 & 185.0  & $6.58 \times 10^{-5}$ \\
$3.66 \times 10^4$ & 1143.1 & 174.2 & 6.56  & $1.63 \times 10^4$ & 0.1832 & 218.9  & $4.41 \times 10^{-5}$ \\
$3.73 \times 10^4$ & 1150.3 & 247.9 & 4.64  & $7.34 \times 10^3$ & 0.2154 & 261.1  & $3.66 \times 10^{-5}$ \\\hline
$3.09 \times 10^4$ & 1079.5 & 124.3 & 8.69  & $3.18 \times 10^4$ & 0.1165 & 207.82 & $3.59 \times 10^{-5}$ \\
$3.09 \times 10^4$ & 1079.5 & 184.8 & 5.84  & $1.24 \times 10^4$ & 0.1802 & 232.2  & $2.67 \times 10^{-5}$ \\
$3.11 \times 10^4$ & 1081.8 & 255.8 & 4.23  & $5.96 \times 10^3$ & 0.1955 & 287.3  & $1.96 \times 10^{-5}$ \\\hline
$2.69 \times 10^4$ & 1031.8 & 110.0 & 9.38  & $3.83 \times 10^4$ & 0.0868 & 232.4  & $2.78 \times 10^{-5}$ \\
$2.69 \times 10^4$ & 1031.8 & 174.8 & 5.90  & $1.27 \times 10^4$ & 0.1614 & 243.4  & $2.43 \times 10^{-5}$ \\
$2.99 \times 10^4$ & 1061.0 & 211.5 & 5.05  & $8.89 \times 10^3$ & 0.1682 & 287.5  & $1.99 \times 10^{-5}$ \\\hline
$2.58 \times 10^4$ & 1017.2 & 76.1  & 13.38 & $9.19 \times 10^4$ & 0.0631 & 212.6  & $2.66 \times 10^{-5}$ \\
$2.58 \times 10^4$ & 1017.2 & 119.3 & 8.53  & $3.03 \times 10^4$ & 0.1321 & 228.1  & $2.20 \times 10^{-5}$ \\
$2.79 \times 10^4$ & 1043.3 & 146.3 & 7.13  & $1.98 \times 10^4$ & 0.1411 & 276.7  & $1.73 \times 10^{-5}$ \\\hline
$2.48 \times 10^4$ & 1003.5 & 125.8 & 7.97  & $2.59 \times 10^4$ & 0.1200 & 193.1  & $4.11 \times 10^{-5}$ \\
$2.48 \times 10^4$ & 1003.5 & 203.6 & 4.93  & $8.41 \times 10^3$ & 0.2041 & 221.8  & $3.03 \times 10^{-5}$ \\
$2.55 \times 10^4$ & 1013.5 & 206.6 & 4.91  & $8.32 \times 10^3$ & 0.1945 & 237.0  & $3.00 \times 10^{-5}$ \\\hline
$1.91 \times 10^4$ & 919.7  & 77.3  & 11.91 & $6.88 \times 10^4$ & 0.0422 & 165.1  & $3.99 \times 10^{-5}$ \\
$1.91 \times 10^4$ & 919.7  & 128.4 & 7.16  & $2.00 \times 10^4$ & 0.1237 & 191.0  & $2.98 \times 10^{-5}$ \\
$1.93 \times 10^4$ & 922.7  & 207.1 & 4.46  & $6.70 \times 10^3$ & 0.2017 & 224.1  & $2.96 \times 10^{-5}$ \\\hline
$1.83 \times 10^4$ & 906.8  & 80.9  & 11.21 & $5.93 \times 10^4$ & 0.0875 & 151.1  & $4.82 \times 10^{-5}$ \\
$1.83 \times 10^4$ & 906.8  & 121.6 & 7.46  & $2.20 \times 10^4$ & 0.1535 & 186.9  & $2.96 \times 10^{-5}$ \\
$1.82 \times 10^4$ & 906.1  & 156.8 & 5.78  & $1.21 \times 10^4$ & 0.1781 & 224.6  & $2.41 \times 10^{-5}$ \\
\bottomrule
\end{tabular}
\end{table*}
\label{lastpage}

\end{document}